Short Paper

# Development of Dynamic Local Area Network (LAN) Based Mock Board Examination System

Charis Ann M. Sancho
College of Computer Studies, Mindoro State College of Agriculture and Technology

Uriel M. Melendres
College of Computer Studies, Mindoro State College of Agriculture and Technology
urielmelendres@minscat.edu.ph
(corresponding author)

## Abstract

*Purpose* – Mock board exam is necessary to identify if the students are ready to take the board exam. However, preparing for the examination is not easy. It takes too much time to release the result, given that the College has limited personnel. Thus, the proponents developed the Dynamic Local Area Network (LAN) Based Mock Board Examination System that makes the preparation and checking easy.

*Method* – The proponents followed the iterative waterfall model to develop the system efficiently. Some criteria of ISO 25010 were adopted in the evaluation instrument. Simultaneously, evaluators are composed of program chairperson, College of Computer Studies (CCS) faculty, board program students, and alumni.

*Results* – The result of 4.91 in Functional Suitability, 4.87 in Performance Efficiency, 4.91 in Usability, 4.90 in Security, and 4.92 in Maintainability shows that the system is fully functional and is usable by any board program.

*Conclusion* – The computer-based examination, implemented through LAN, can simplify administering personnel of MinSCAT mock board examination. The development of the "Dynamic LAN Based Mock Board Examination System" is a great help to the MinSCAT and their board program graduates if it is implemented.

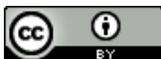

*Recommendations* – The system has hidden weaknesses that were only identified in actual operation. Thus, further testing is needed, like beta testing, to identify and correct it for better performance. After beta-testing, the system can be improved through iteration of the Waterfall Model phases.

*Research Implications* – The developed system can improve the mock board examination process and ease preparing this significant examination. Moreover, it immediately provides a result that helps the examinees to identify their weaknesses and do a further review to master it.

## INTRODUCTION

The examination is to assess the knowledge of a person in a particular subject or field. Undoubtedly, examination played a vital role in our lives, especially in education. Most of the professional programs offered in the Universities and Colleges have licensure or board exams. A board examination in the Philippines performed through a written test in different core topics of particular studies. Passing the board exam qualifies the graduates to practice their profession or able to work in the government. Relative to this, many review centers emerged that conducting a comprehensive review of specific board exams. Aside from those review centers, many Colleges and Universities offered a local review class for their graduates to pass.

Mindoro State College of Agriculture and Technology (MinSCAT) is a state college in the province of Oriental Mindoro, Philippines, that offers quality but affordable education to the youth. MinSCAT has three campuses, and each campus offered several board programs. Bongabong Campus offered four Board programs while the Calapan City Campus has three, and Main Campus also has four. To cater to their graduates' needs in respective board programs, MinSCAT also offered a review and mentoring classes before and after graduation. At the end of the review, they were conducting a mock board, which is essential to determine if graduates are ready to take the actual exam. The study conducted by Gono (2018), shows that mock board result is an effective medium to predict if the examinees could pass the board exam or not. This concept is supported by another study, which showed that the mock board examination result is a valid predictor of the graduates' performance in licensure exams (John & Ercia, 2017).

Traditional paper and pencil tests were used in the actual board examination. Still, there are many benefits if the mock board can be done using modern methods. Many institutions are now using a computerized examination system, which avoids unnecessary burdens of administering regular examinations. Also, regular assessment is prone to human errors (Abass et al., 2017). The exam preparation could take weeks or months, and the examinees need to wait for a long time to release the result. Limited administering personnel is the primary reason for the delay. It is crucial to announce the result immediately to know the examinee's weaknesses. The result could help examinees perform further studies on the topics with low scores to improve their performance (Meyerson et al., 2017). The result is more reliable because the proctored computerized exam is more resistant to cheating than those not proctored (Daffin & Jones, 2018).

Many learning institutions used Learning Management System (LMS) for managing their educational content, including examinations. Most of the emerging LMS nowadays has an excellent assessment tool (Khramtsova & Olentsova, 2019). But most of the LMS assessment tools are designed for quizzes or short examination only since its primary purpose is managing learning contents. A system intended only for examination is better when administering an extended test like mock board exams.

The information above helps the proponents to conceptualized a web-based mock board examination system. The study aims to design and develop an Examination Management system specialized only for mock board exams. The system was entitled "Dynamic Local Area Network (LAN) Based Mock Board Examination System." The system provides MinSCAT with a manageable interface in creating and administering mock board examinations. Also, to reduce the time and effort in conducting the mock board exams since the usual paper and pencil tests are very time-consuming to prepare and check. It also eliminates the use of the bulk of documents, which is good for the environment.

The developed system can run in a local area network (LAN), whether wired or wireless. The course is a web-based developed through CodeIgniter, a PHP framework, and other client-side scripting languages. The system interface was designed to be user-friendly and dynamic, making it suited for any board programs of the College.

## LITERATURE REVIEW

Modern technology improved almost all aspects of our education system, including learner's assessment. With a commercial and open-source Learning Management System (LMS), many learning institutions efficiently assess their students learning through embedded modules for short quizzes and examinations. In this part, the proponents present characteristics and features of five known open-source LMS specifically on their quiz tool.

Obeidallah and Shdaifat (2020) evaluated the assessment tool of the ten most known open-source LMS. According to the result, Moodle, Dokeos, ATutor, Sakai, and Opigno have the best assessment tool. These five identified LMS was also used to compare to the developed system. Table 1 shows the comparison of the relevant features of those mentioned LMS and the developed system.

Table 1. Comparison of LMS (Obeidallah and Shdaifat, 2020) and the Developed System

| LMS Features | Moodle | Dokeos | ATutor | Sakai | Opigno | Developed System |
|---|---|---|---|---|---|---|
| Item Analysis | Yes | Yes | Yes | Yes | Yes | Yes |
| Randomized Questions | Yes | Yes | Yes | Yes | Yes | Yes |
| Question Categorization | Yes | Yes | Yes | Yes | Yes | Yes |
| Question banks | Yes | Yes | Yes | Yes | Yes | Yes |
| Grade Report | Yes | Yes | Yes | Yes | Yes | Yes |
| Time Limit | Yes | Yes | Yes | Yes | Yes | Yes |
| Implementation | Online | Online | Online | Online | Online | Offline |

The salient features of five LMS assessment tools are also present in the developed system, as shown in Table 1. The only difference is that all LMS is implemented online, which requires an internet connection while the developed system is offline. Furthermore, LMS is designed to facilitate teaching and learning activities among teachers and students and not an examination management system, although assessment tools are an essential feature (Kasim & Khalid, 2016; Anand & Eswaran, 2018). On the other hand, the developed system was dedicated only to mock board examinations, with no unnecessary features that make it lighter and easy to use.

## METHODOLOGY

### *Software Development Method*

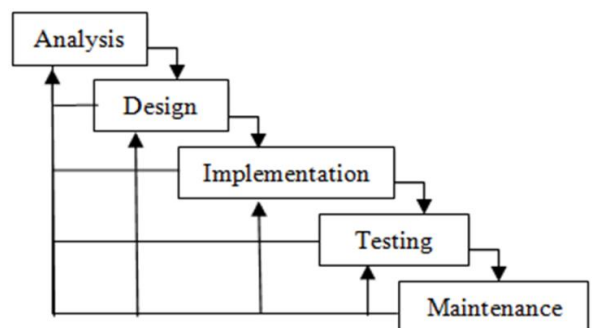

*Figure 1.* Iterative Waterfall Model (Soni & Agal, 2017)

There are many choices when choosing a software development model for developing software solutions. The traditional waterfall model was still widely used, given that various emerging models have impressive outcomes (Almeida & Simões, 2019). Maybe, because it is simple, easy to understand, and allows the proponents to identify specific deliverables in every phase. But the most critical weakness of this traditional method is that it does not let go back to the previous stage to handle changes (Egwoh & Nonyelum, 2017). This weakness leads to developing a new method known as the iterative waterfall model (Figure 1). The two's characteristics are almost the same except, the iterative model allows iteration with the preceding and succeeding steps to enable user participation and satisfy the feedback (Van Casteren, 2017). There are also other modern methods, but most applicable for large and complex software projects (Singh & Kaur, 2019).

The proposed system falls in small-scale projects, less complicated, and the requirements are easy to specify. For these reasons, the iterative waterfall model is appropriate to use. Other than that, this method is flexible and straightforward, although limitations are inevitable (Misra et al., 2016). Since the model allows feedback, the proponents tend to repeat each stage several times, making it uneasy to manage and time-consuming. Despite the limitations, the guarantee of success rate is high compared to the traditional and other existing models (Van Casteren, 2017). Furthermore, the proposed system is not that complex to have a severe management problem.

Iterative Waterfall phases are strictly followed to finish the system on the set period and ensure its quality. Table 2 shows different activities conducted in every phase.

Table 2. Activities in Every Phase

| Phases | Salient Activities |
| --- | --- |
| Analysis | Finalization of user requirements, functional, non-functional, and other conditions that are essential in the development. Identification of different software development tools like scripting languages, databases, and other third-party software helps develop the system efficiently. |
| Design | Creating different diagrams could help design like a network diagram, use case, and data flow diagram. The user interface is also visualized with a simple prototype. |
| Implementation | The proposed system design is implemented through coding using different development tools and programming languages. Some design discrepancy was identified and not able to implement for some reason. Since the process is iterative, proponents need to go back to the previous phase to slightly adjust the system design. The system was divided into small units called modules for every identified features and functionality. All working modules were integrated into the primary system and checked its compatibility. |
| Testing | After coding, the system underwent alpha testing to ensure that the system is functional. The system was evaluated using the identified criteria of ISO 25010. Some flaws are identified and need to fix; thus, going back to the implementation phase is required. |
| Maintenance | Not yet perform because the system is not, however, fully implemented. |

## System Design

The system is consisting of two primary modules, such as examinee and admin. The examinee module is the webpage that can be accessed by the mock board takers. In contrast, the admin module can only access by managing personnel.

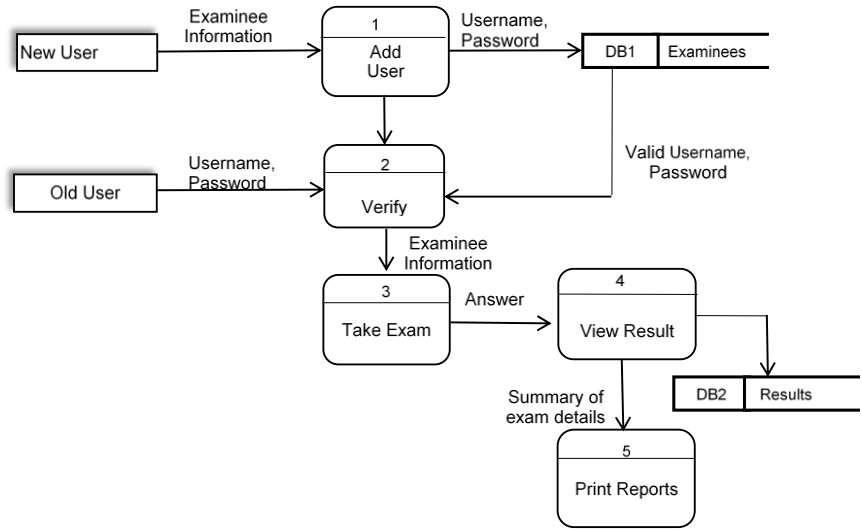

*Figure 2.* Data Flow Diagram (DFD) of the Examinee Module

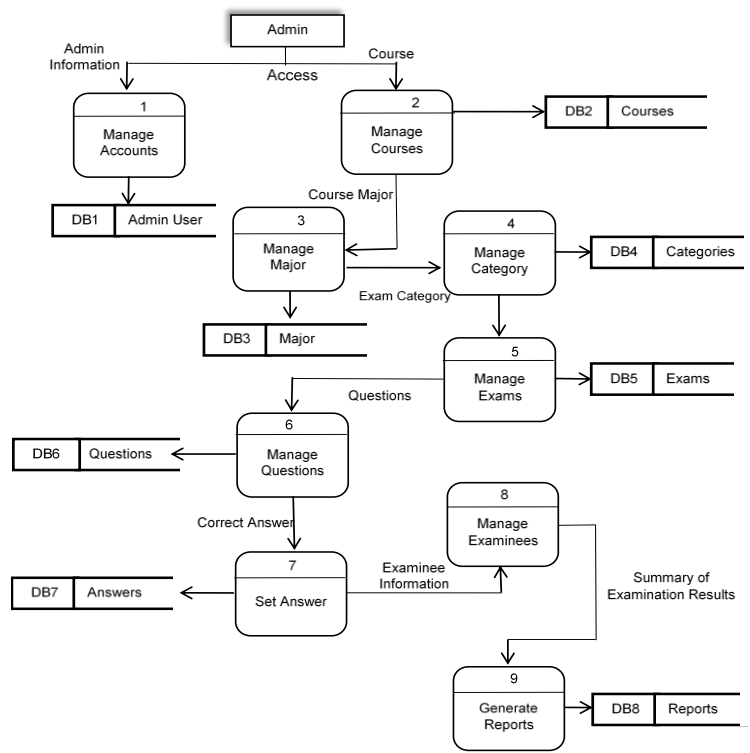

*Figure 3.* DFD of the Admin Module

Figures 2 and 3 show the data flow diagrams of two modules. The only information that the examinee could access is a particular examination questionnaire and his/ her examination result. Everything can be managed on the admin side, like adding and verifying users, examinations, and printing reports.

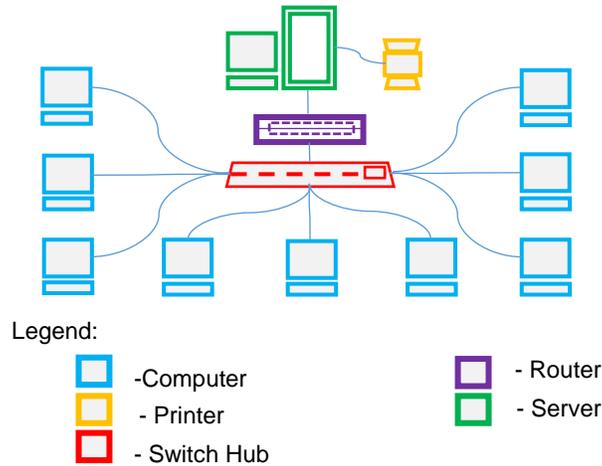

Legend:
- Computer
- Printer
- Switch Hub
- Router
- Server

*Figure 4.* Network Diagram

The system implementation is through LAN. The network design (Figure 4) is straightforward and lacks expensive networking devices that make the system feasible for implementation. There is one server computer where the printer is connected and enough client computers to accommodate examiners. The users need to be connected to the router to obtain the IP Address required by the clients. The localhost server must be opened to access the web page by the client computers. On the client-side, it is necessary to type the correct IP Address of the server and the system's name, e.g., 192.168.100.10/mockboard/login/dean.

## Hardware and Software Requirements

The system can run on any platform since it was accessible through a web browser like Google Chrome. Table 3 and Table 4 provided the hardware and software used in the development of the proposed system.

Table 3. Software Requirement

| Software | Specification |
| --- | --- |
| Operating System | Windows 10 |
| Localhost Web server Solution | XAMPP 3.2.2 |
| Database | MySQL |
| Browser | Google Chrome |
| Mark-up, Programming Language and Framework | HTML, CSS, PHP, JavaScript, CodeIgniter |
| Text Editor | Notepad++, SublimeText |

Table 4. Hardware Requirements

| Device | Specification |
|---|---|
| Laptop | Processor: Intel Core i3 2.3GHz |
|  | RAM:4GB |
|  | System Type: 64bit |
| Network Hub | 16 ports |
| Router | Full Duplex |

## *Testing and Evaluation*

Upon initial development, alpha testing was performed by the proponents. The proponents devised a simple alpha testing instrument (Table 5) focused on two primary criteria, such as *Design and Compatibility* and *Features and Functionality*. To ensure the system's efficiency, BS Information Technology Instructors and external expert was invited to participate in the testing.

Table 5. Alpha Testing Instrument

| Criteria | Developer's Remarks |
|---|---|
| Design and Compatibility |  |
|    Browser Compatibility |  |
|    Navigation |  |
|    Display |  |
|    Database Design |  |
| Features and Functionality |  |
|    Log-in/ Register Module |  |
| Examinee Module |  |
|    Admin Module |  |
|    Course Module |  |
|    Examination Module |  |
|    Report Module |  |

For system evaluation, the proponents adopt five criteria of ISO 25010, such as Functional Suitability, Performance Efficiency, Usability, Security, and Maintainability. These criteria can help to know if the system is usable and met the target output. Table 6 shows the devised instrument based on the mentioned criteria. The instrument validation is done by the College Research Director and other faculty researchers of MinSCAT. A five-point Likert scale has been utilized with the mean range interpretation of (1) 1.0-1.80: poor, (2) 1.81-2.60: Fair, (3) 2.61-3.40: Good, (4) 3.41-4.20: Very Good, (5) 4.21-5.00: Excellent.

Table 6. Evaluation Instrument

| Criteria | Indicator |
|---|---|
| Functional Suitability | The system functions help to accomplish user-specified tasks. |
| | The system questions are relevant and show randomly. |
| | The system examination provides accurate results. |
| Performance Efficiency | The system process immediately responds whenever it handles vast data. |
| | The system features and functions work correctly. |
| | The information provided by the system correct. |
| Usability | The system enables the user to use the system with minimal supervision. |
| | The system is interactive with a pleasing interface. |
| | The contents of the system are appropriate for user needs. |
| Security | The system provides security mechanisms for both admin and examinees. |
| | The system prevents unauthorized access to computer programs or data stored on the system. |
| | The system does not compromise the personal information of the user. |
| Maintainability | The system is dynamic and applicable to any board program. |
| | The system is easy to modify without a significant effect on the other components of the system. |
| | The system is maintainable and can be enhanced easily. |

The proponents purposively selected forty (40) evaluators to make the evaluation more reliable. The evaluators are composed of program chairperson, BSIT faculty, board program's faculty, students, and alumni.

## RESULTS

### *Presentation of the System Output*

All users must enter the correct username and password (Figure 5) to access their respective dashboards or click the *register here* to access the registration form (Figure 6). The examinee must fill all the text fields and check the box to agree with the system's terms and conditions. Examinees must select their respective programs or course with their major if applicable.

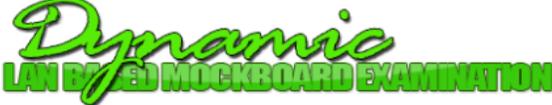

*Figure 5.* Login Form

*Figure 6.* Examinee Registration Form

*Figure 7.* Examinee Dashboard

*Figure 8.* Examination Result

Figure 7 shows the examination details, such as exam name, time limit, passing rate, examination date, total questions, and status of the examinee in every category of the exam. When the examinee is ready, just click the "*Take Exam*" button. It displays the examination page, and the time automatically starts. After taking the exam, the examination result (Figure 8) show in less than a minute.

On the other hand, upon entering the username and password correctly, the system administrator could access the admin dashboard (Figure 9). The dashboard provides a short description of the system and some instruction in managing student and examination. Also, it provides the necessary navigation and settings to create a test. Figure 10 shows the list of registered examinees and pending applicants. The admin must verify all examinees to login.

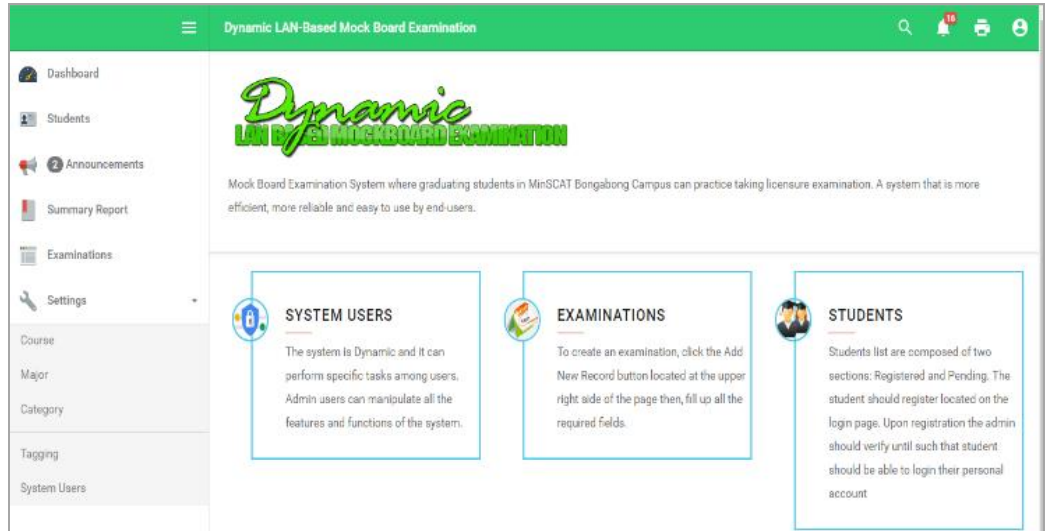

*Figure 9.* Admin Dashboard

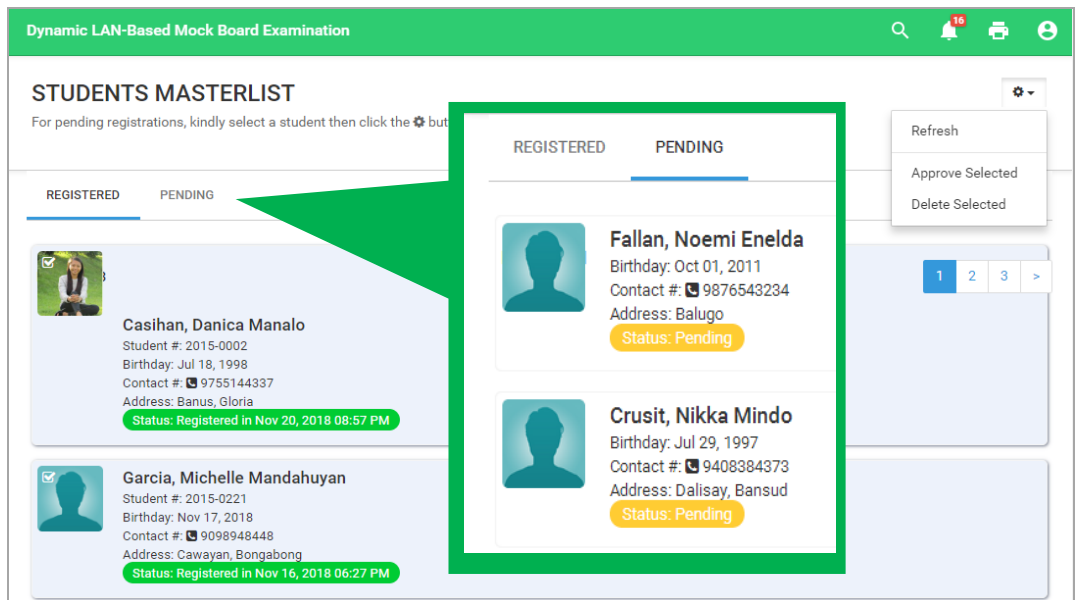

*Figure 10.* Examinees' List

*Figure 11.* Adding Course

The admin can also add or delete programs or courses and its corresponding specialization (Figure 11). This feature allows to manage different programs and avoid mixing of questionnaires. When creating an examination, the administrator must click the setting icon to access the *adding exam form* (Figure 12). Figures 14-15 shows in detail how to input questions, choices, and set up the correct answer.

*Figure 12.* Creating Examination

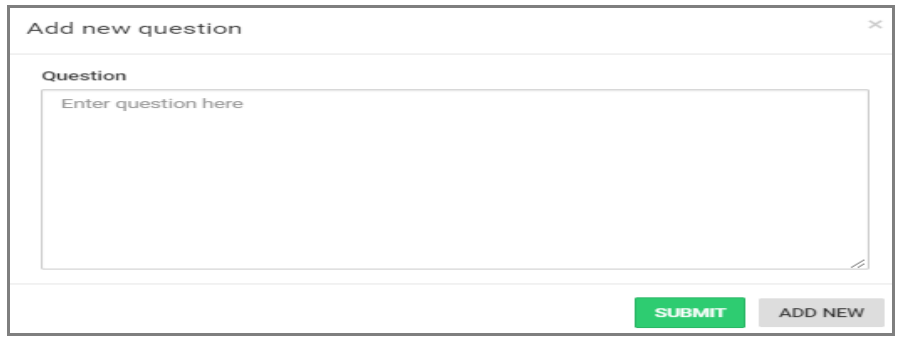
*Figure 13.* Inputting Questions

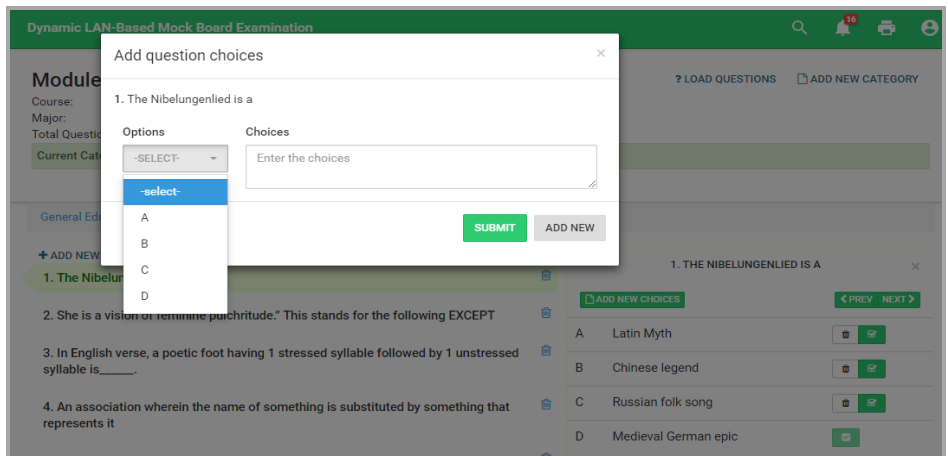
*Figure 14.* Adding Choices

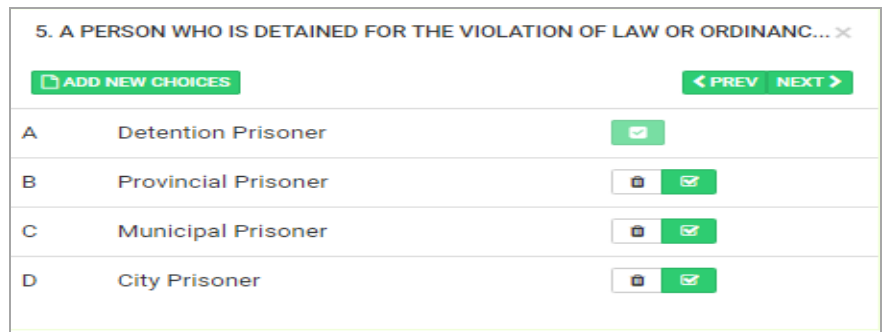
*Figure 15.* Setting Correct Answer

## Testing and Evaluation Result

The system underwent beta testing to ensure the quality and efficiency of the Mock Board Examination System. Table 7 shows the beta testing result. On the *Design and Compatibility* criteria, minor problems were noticed on compatibility view even though it cannot affect the overall functionality, but it essential to fix. The proponents validated the codes and adjusted some CSS source codes. Besides, idle database tables were dropped to enhance the loading speed of the system. Whereas, many problems and errors were discovered in *Features and Functionality*. Most of the issues need exhaustive debugging, which takes weeks to finish. After fixing all errors and design issues identified in alpha testing, the system underwent system evaluation with chosen respondents.

Table 7. Beta Testing Result

| Criteria | Remarks |
|---|---|
| Design and Compatibility | |
| Browser Compatibility | Running in four commonly (Chrome, Firefox, Opera, Internet Explorer (IE)) used browsers. Still, there are minor display issues in Opera and IE. |
| Navigation | The navigation is concise and clear. |
| Display | Shows consistency, Font-size, style, and colors are appropriate. Good visual contrast. |
| Database Design | Some tables are not used and need to delete. |
| Features and Functionality | |
| Log-in/ Register Module | Log-in/register is functional, but there is some unnecessary examinee information. |
| Examinee Module | Cannot access the examination page |
| Admin Module | The system allows many admins, but the content is similar. There is no distinction between different program admins. |
| Course Module | There is some error when adding subjects of programs with majors like BSEd. |
| Examination Module | We are encountering errors when adding a questionnaire and its choices. |
| Report Module | This part is not responding because of examination module issues. |

The system was evaluated by the Program Chairperson of BSEd, BSFi, and BSCrim, 4 BSIT faculty, three other faculty teaching in BSEd and BSCrim, 30 graduating students from BSFi, BSEd, and BSCrim, and five board program alumni. The average mean in all criteria was interpreted as excellent, as shown in Table 8, which indicates that the developed system is useful and feasible for implementation. Also, notice that the variance is less than one, which revealed that the respondents' rating is adjacent. Meaning almost all respondents agreed that the system is excellent. It is also reflected that the system is dynamic and can respond to the target users' needs.

Table 8. Summary of System Evaluation Result

| Criteria | Average Mean | Variance |
|---|---|---|
| Functional Suitability | 4.91 | 0.0963 |
| Performance Efficiency | 4.87 | 0.1117 |
| Usability | 4.91 | 0.0803 |
| Security | 4.90 | 0.0932 |
| Maintainability | 4.92 | 0.0897 |

## DISCUSSION

The use of computer-based examination provides many benefits to the administering personnel. This method reduced the time allotted for preparation, scoring, result analysis, and ease of administration (Debuse & Lawley, 2016). Computer networking also plays a vital role in the implementation of computerized systems. Although the use of the internet in the computerized examination is noticeable, LAN is the most appropriate and convenient because the mock board examination needs rigid supervision to avoid cheating. While the computerizing examination is not new, and many LMS already have this feature. Still, developing a new examination system specialized only for the mock board examination is essential for MinSCAT mock board administering personnel.

The developed system provides an easy to operate user interface and full of relevant features present in a sound examination management system. The evaluation result proved that the user requirements' target requirements were achieved, which denotes that the system can be implemented.

## CONCLUSIONS AND RECOMMENDATIONS

The computer-based examination, implemented through LAN, can simplify administering personnel of MinSCAT mock board examination. The development of the "Dynamic LAN Based Mock Board Examination System" is a great help to the MinSCAT and their board program graduates if it is implemented. Since it is dynamic, every board program of MinSCAT can easily use it. The system evaluation results in five criteria of ISO 25010 are all interpreted as excellent, which shows that the system is functional, efficient, and feasible for implementation. Nevertheless, there are other aspects to consider before the implementation of the system. Thus, the proponents suggest conducting beta testing to identify some problems that could occur in actual operation. The beta testing results could help to improve the system further.

Compared to widely used examination software, the developed system has more to improve. Through collaboration with the target users and further analysis of existing systems, many features can be added. For example, machine learning can be integrated into the system to determine whether the questionnaire is "good" or "bad." It may optimize the student scores that could help to improved student assessment. The iterative process of the waterfall can be implemented for the proposed improvements.

## IMPLICATIONS

The evaluation of the developed system for mock board examination has an excellent result. Thus, once the system was implemented can improve the mock board examination process and ease the burden of preparing this significant examination. Moreover, it immediately provides a result that helps the examinees to identify their weaknesses and do a further review to master it.

# ACKNOWLEDGEMENT

The proponents would like to thank the Research Department of MinSCAT Bongabong and the whole College of Computer Studies family that encourages us to finish and publish this study. Also, we would like to thank the Program Chairperson and other faculty members of board programs of MinSCAT for their cooperation during the conduct of the study.